\documentstyle[12pt,epsfig]{article}

\def\beq{ \begin{equation}}
\def\eeq{\end{equation} }
\def\bea{\begin{eqnarray}}
\def\eea{\end{eqnarray}}

\def\ohsq{\Omega_{\chi} h^2}

\def\eps{\epsilon}

\def\ra{\rightarrow}

\def\bar{\overline}

\def\gappeq{\mathrel{\raise.3ex\hbox{$>$\kern-.75em\lower1ex\hbox{$\sim$}}}}
\def\lappeq{\mathrel{\raise.3ex\hbox{$<$\kern-.75em\lower1ex\hbox{$\sim$}}}}

\begin{document}


\title{\begin{flushright}
\vspace*{-1.2 cm}
{\normalsize hep-ph/0103256} \\
\vspace*{-0.4 cm}
{\normalsize CERN-TH/2001-085}  \\
\vspace*{-0.4 cm}
{\normalsize FISIST/06-2001/CFIF}  \\
\end{flushright} 
\vspace*{0.7 cm} {\Large \bf Charged-Lepton-Flavour Violation in the CMSSM
in View of the Muon Anomalous Magnetic Moment}
\author{{\normalsize 
\bf D.~F. Carvalho$^a$, John Ellis$^b$, M.~E. G\'omez$^a$}
and {\normalsize \bf S. Lola$^b$ }\\
{\small a) CFIF, Departamento de Fisica, Instituto Superior T\'ecnico,
Av. Rovisco Pais,} \\
{\small 1049-001~Lisboa, Portugal} \\
{\small b) CERN, CH-1211 Geneva 23, Switzerland } \\ 
}}
\date{}
\maketitle
\vspace*{-0.5 cm}
{\bf Abstract:
{\small
 
We use the BNL E821 measurement of $g_\mu - 2$, the anomalous magnetic
moment of the
muon, to normalize, within a supersymmetric GUT framework,
constrained MSSM (CMSSM)  predictions for processes that violate
charged-lepton flavour conservation, including $\mu \rightarrow e \gamma$,
$\mu \ra e$ conversion and $K^0_L \rightarrow \mu^\pm e^\mp$. We
illustrate our analysis with two examples of lepton mass matrix textures
motivated by data on neutrino oscillations. We find that $\mu \rightarrow
e \gamma$ may well occur at a rate within one or two (two or three) orders
of magnitude of the present experimental upper limit if $g_\mu - 2$ is
within the one- (two-)standard deviation range indicated by E821. We also
find that $\mu \ra e$ conversion is likely to occur at rate measurable by
MECO, and there is a chance that $K^0_L \rightarrow \mu^\pm e^\mp$ may be
observable in an experiment using an intense proton source. 

}}

\begin{flushleft}
\vspace*{1.2 cm}   
{\normalsize CERN-TH/2001-085}  \\
{\normalsize FISIST/06-2001/CFIF}  \\
{\normalsize March 2001} \\
\end{flushleft}

\newpage

\section{Theoretical Framework}

The observation of neutrino oscillations \cite{bks,SKam}
implies that the individual
lepton numbers $L_{e, \mu, \tau}$ are violated, suggesting the appearance
of processes that violate charged-lepton-number, 
such as $\mu \rightarrow e
\gamma$, $\mu \rightarrow 3 e$, $\mu \ra e$ conversion on heavy nuclei
\cite{KO}, $\tau
\rightarrow \mu \gamma$ \cite{GELLN} and $K^0_L \ra \mu e$~\cite{KAONS}. 
The present experimental upper limits on these processes are ${\cal B}(\mu
\ra e \gamma) < 1.2 \times 10^{-11}$~\cite{Brooks}, ${\cal B}(\mu^+ \ra
e^+ e^+ e^-) < 1.0 \times 10^{-12}$~\cite{Bellgardt}, ${\cal B}(\mu^- Ti
\ra e^- Ti) < 6.1 \times 10^{-13}$~\cite{Wintz}, ${\cal B}(\tau \ra \mu
\gamma) < 1.1 \times 10^{-6}$~\cite{CLEO} and ${\cal B}(K^0_L \ra
\mu^\pm e^\mp) <
4.7 \times 10^{-12}$~\cite{kaon1}. 
On the other hand, in minimal GUT models where the small
neutrino masses are generated by the see-saw mechanism with massive
singlet neutrinos $\nu_R$, and there are no new lighter particles, the
amplitudes for charged-lepton-flavour violation are proportional to
inverse powers of the heavy singlet neutrino mass $M_{\nu_R}$, and the
rates for rare decays are extremely suppressed~\cite{neutrinoLFVns}. 

However, the observation of an apparent discrepancy between the measured
value of
the anomalous magnetic moment of the muon, $a_\mu \equiv (g_\mu - 2)/2$,
and the value predicted in the Standard Model 
\cite{BNLE821} suggests the appearance of
new physics at the TeV scale in the lepton sector, with superseymmetry
being one of the favoured options~\cite{ENO,othersusy}.
Moreover, there is a striking resemblance between the effective operators
that generate
$\mu\rightarrow e \gamma$ and $\delta a_\mu$:
\begin{eqnarray}
{\cal L}_{eff} = 
e \frac{m_{\ell_j}}{2}\bar{\ell}_i \sigma_{\mu\nu}  
F^{\mu\nu} (A_M^{L,ij} P_L+A_M^{R,ij} P_R) \ell_j
\label{eff_op}
\end{eqnarray}
resulting in
\begin{eqnarray}
{\rm Br}(\mu \rightarrow e \gamma)  = 
\frac{48\pi^3 \alpha}{G_F^2} (|A_M^{L,12}|^2+|A_M^{R,12}|^2)
\end{eqnarray}
and
\begin{eqnarray}
\delta a_\mu=\frac{m_\mu^2}{2} (A_M^{L,22}+A_M^{R,22}).
\end{eqnarray}
Hence, if these quantities are dominated by either the
$A_M^L$ or the $A_M^R$, there is a direct relation between them:
\begin{equation}
{\cal B}(\mu \ra e \gamma) = {192 \pi^3 \alpha \over G_F^2 m_\mu^4}
\times (\delta a_\mu)^2 \times \epsilon^2,
\label{relation}
\end{equation}
where the lepton mixing factor $\epsilon \equiv A_M^{L/R,12}/
A_M^{L/R,22}$. We see explicitly from (\ref{relation}) that the 
apparent measurement of $\delta a_\mu$ enables the rate
for $\mu \ra e \gamma$ to be predicted, in the
context of any model of lepton flavour violation motivated by the
observations of neutrino oscillations which is able to predict
$\epsilon$.

\begin{figure}[h]
\begin{center}
\epsfig{file=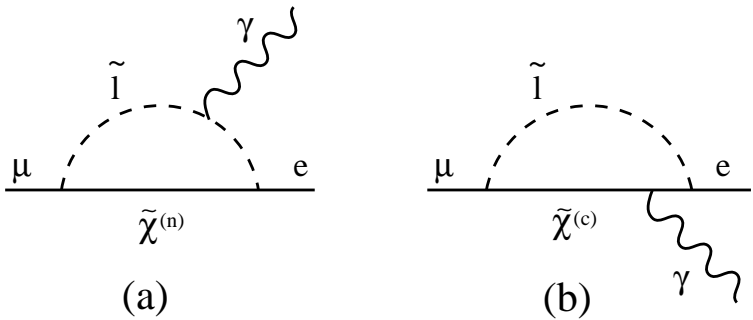,width=10cm}
\end{center}
\caption{{\it Generic Feynman diagrams for $\mu\ra e\gamma$
decay: $\tilde l$ represents a charged slepton (a) or
sneutrino (b), and
$\tilde\chi ^{(n)}$ and $\tilde\chi ^{(c)}$ represent  
neutralinos and charginos respectively.}}
\label{figure1}
\end{figure}

One example of a theory where this connection can be made is
supersymmetry~\cite{HT,othermueg}. In a supersymmetric model, the
amplitudes for
processes
violating charged lepton number are suppressed by inverse powers of the
supersymmetry-breaking scale, which is thought to be at most $1$~TeV.
In particular, in the presence of
$\tilde{\mu}$-$\tilde{e}$ ($\tilde{\nu}_\mu$-$\tilde{\nu}_{e}$)
mixing,
the diagrams of Fig.~\ref{figure1} are generated, which are isomorphic to
the corresponding flavour-conserving diagrams contributing to $\delta
a_\mu$. If the dominant supersymmetric contribution to $g_\mu-2$ comes
from
the chargino-sneutrino diagram Fig.~\ref{figure1}(b) involving left-handed
leptons, one expects the chargino diagrams also to dominate
$\mu\rightarrow e \gamma$~\footnote{When the right-handed sleptons also
contribute significantly to the lepton-flavour-violating masses, the
neutralino--slepton diagrams of Fig.~\ref{figure1}(a)  also contribute to
$\mu\rightarrow e \gamma$, modifying the correlation between
$\mu\rightarrow e \gamma$ and $g_\mu-2$.}.  Taking the BNL E821
measurement of the muon anomalous magnetic moment~\cite{BNLE821} at its
face value fixes
the overall mass scale of the sparticles circulating in the loops in
Fig.~\ref{figure1}, and a supersymmetric GUT model of $\nu_\mu$-$\nu_e$
mixing
can be used to calculate the amount of $\tilde{\mu}$-$\tilde{e}$ and
$\tilde{\nu}_\mu$-$\tilde{\nu}_{e}$ mixing, i.e., the factor $\epsilon$ in
(\ref{relation}), enabling the rate for $\mu \ra e \gamma$ to be
predicted.

Connections to other processes violating charged-lepton flavour, such
as $\mu \ra 3e$ and $\mu \rightarrow e$ conversion on nuclei, can be made
in a similar way, though less directly.  Among these processes, $\mu \ra 
e$
conversion is particularly promising, since the present experimental
sensitivity may be improved by several orders of magnitude in future
experiments such as MECO~\cite{MECO} and PRISM~\cite{KO}.  
These processes receive important contributions
from photon
and $Z$ `penguin' diagrams, which are related to those for $\mu \ra e
\gamma$ via a virtual gauge boson coupling to an $e^+e^-$ or a
quark-antiquark pair, but also from box diagrams and their supersymmetric
analogues.  The dominant photonic contribution yields~\cite{KO}
${\cal B}({\mu
Ti\ra e Ti})\approx 5.6 \times 10^{-3} {\cal B}(\mu\ra e\gamma)$, but this
ratio may
receive substantial corrections from subdominant contributions, which we
take into account in our calculations.  Finally, we also mention the
possibility in supersymmetric theories of observing charged-lepton-flavour
violation in $K^0_L \ra \mu e$ decay~\cite{KAONS}. This involves the
mixing of squarks as well as that of sleptons, thus providing additional
information. The rate is small in the CMSSM with
right-handed neutrinos, but might be observable
using a future intense proton source, if $\tan\beta$ is
large~\footnote{Larger rates are also possible in supersymmetric models
with broken $R$ parity~\cite{KAONS}, a possibility not considered in
this paper.}. 

In all these processes, the magnitudes of the rates depend on the masses
and mixings of sparticles.  Excessive rates for charged-lepton-flavour
violation are generically predicted in models with non-universal scalar
masses at the GUT scale.  Thus we consider constrained MSSM (CMSSM) 
models that respect this universality, such as minimal
supergravity (mSUGRA)~\cite{mSUGRA},
gauge-mediated supersymmetry ~\cite{gaugmed} and no-scale
models~\cite{Ellis:1984bm}.  In such models, even though there are no
off-diagonal contributions to the sfermion mass matrix at ${M_{GUT}}$,
renormalization effects due to lepton Dirac Yukawa couplings within the
Minimal Supersymmetric Standard Model (MSSM) with massive neutrinos spoil 
this diagonal form ~\cite{neutrinoLFVs}, making these processes
observable. 

In this paper, we calculate these quantum corrections
in the context of the most natural
mechanism for obtaining sub-eV neutrino masses, namely
the see-saw mechanism~\cite{seesaw}. This
involves Dirac neutrino masses $m_{\nu_D}$ of the same order as
the charged-lepton and quark masses, and heavy Majorana
masses $M_{\nu_R}$, leading to a light effective neutrino mass
matrix:
\begin{equation}
m_{eff}
=m_{\nu_D}\cdot (M_{\nu_R})^{-1}\cdot m^{{\normalsize T}}_{\nu_D}.
\label{eq:meff}
\end{equation}
Neutrino-flavour mixing may then occur through either the
Dirac and/or the Majorana mass matrices, which may also
feed flavour violation through to the charged leptons.

In general, the Dirac neutrino and charged-lepton Yukawa
couplings,  $\lambda_{\nu_D}$ and $\lambda_{\ell}$ respectively,
cannot be diagonalized simultaneously.
Since both these sets of lepton Yukawa
couplings appear 
in the renormalization-group equations,
the lepton Yukawa
matrices and the slepton mass matrices cannot be 
diagonalized simultaneously at low energies, either.
In the basis where $\lambda_{\ell}$ and  $m_{\ell}$
are diagonal, the slepton-mass matrix acquires 
non-diagonal contributions from renormalization at scales below
$M_{GUT}$, of the form:
\bea
\delta{m}_{\tilde{\ell}}^2\propto \frac 1{16\pi^2} (3 + a^2)
\ln\frac{M_{GUT}}{M_N}\lambda_{\nu_D}^{\dagger} 
\lambda_{\nu_D} m_{3/2}^2,
\label{offdiagonal}
\eea
where $a$ is  related to the trilinear mass parameter:
$A_\ell \equiv a m_{0}$, where
$m_0$ is the common assumed value of the scalar masses at the GUT
scale. 

Different oscillation scenarios for the atmospheric and solar neutrino
deficits~\cite{rev}, e.g., those with small/large neutrino
mixing angles and with eV or much lighter neutrinos, predict in general
different rates for lepton-flavour violation.  Typically, the larger the
$\nu_\mu$-$\nu_e$ mixing and the larger the neutrino mass scale, the
larger the rates. Thus, models of degenerate neutrinos with bimaximal
mixing lead to significantly larger effects than, for instance,
hierarchical neutrinos with a small vacuum mixing angle. Just-so
vacuum solutions to the solar neutrino 
deficit with $\delta m^2 \approx
10^{-10}$ eV$^2$ typically predict small rates if the neutrino masses are
hierarchical, even if the (1-2) mixing angle is large. 

\section{Sample Models of Neutrino Masses and Mixing}

In order to illustrate our estimates of the expected effects, we calculate
the rates for rare processes violating charged-lepton number in
representatives of two different types of models, 
one with small  and one
with large $\mu$-$e$ mixing.
The first model (A) is based on
Abelian flavour symmetries and symmetric fermion mass matrices~\cite{IR},
and leads to the following pattern of charged-lepton masses $m_\ell$,
neutrino Dirac masses $m_{\nu_D}$, charged-lepton mixing $V_\ell$ and
Dirac mixing $V_{\nu_D}$ \cite{LLR}:
\bea
m_{\ell }  \propto  \left( 
\begin{array}{ccc}
\bar{\epsilon}^{7} & \bar{\epsilon}^{3} & \bar{\epsilon}^{7/2} \\ 
\bar{\epsilon}^{3} & \bar{\epsilon} & \bar{\epsilon}^{1/2} \\ 
\bar{\epsilon}^{7/2} & \bar{\epsilon}^{1/2} & 1
\end{array}
\right),
m_{\nu_D} \propto \left( 
\begin{array}{ccc}
\bar{\epsilon}^{14} & \bar{\epsilon}^{6} & \bar{\epsilon}^{7} \\ 
\bar{\epsilon}^{6} & \bar{\epsilon}^2 & \bar{\epsilon} \\ 
\bar{\epsilon}^{7} & \bar{\epsilon} & 1
\end{array}
\right),
\eea
\bea
V_\ell = \left(
\begin{array}{ccc}
1 & \bar{\epsilon}^{2} & -\bar{\epsilon}^{7/2} \\
-\bar{\epsilon}^{2} & 1 & \bar{\epsilon}^{1/2} \\
\bar{\epsilon}^{7/2} & -\bar{\epsilon}^{1/2} & 1
\end{array}
\right), V_{\nu_D} = \left(
\begin{array}{ccc}
1 & \bar{\epsilon}^{4} & -\bar{\epsilon}^{7} \\
-\bar{\epsilon}^{4} & 1 & \bar{\epsilon} \\
\bar{\epsilon}^{7} & -\bar{\epsilon} & 1
\end{array}
\right), \label{Asolutions}
\eea
where $\bar{\epsilon}$ is a (small) expansion parameter
related to the Abelian symmetry-breaking scale.
In this model, even a charged-lepton matrix with large (2-3) mixing always
predicts small $\mu-e$ mixing, as a result of fixing the charged-lepton
mass hierarchies~\cite{LLR}. 

As a second possible model (B), we discuss the case of bimaximal
mixing
appearing in \cite{bar}.
In the basis where the charged-lepton mass matrix is
diagonal, the neutrino mixing matrix is:
\beq
V_{\nu_D}
= \left( \begin{array}{ccc}
 {1\over\sqrt2} & -{1\over\sqrt2} & 0 \\
 {1\over2}      &       {1\over2} & -{1\over\sqrt2} \\
 {1\over2}      &       {1\over2} & {1\over\sqrt2} \\
\end{array} \right)
\label{U}
\eeq
corresponding to a neutrino mass matrix of the form
\beq
m_{eff} = m \left[
\left( \begin{array}{ccc}
0 &  0 & 0 \\
0 &  1 & -1 \\
0 & -1 & 1 \\
\end{array} \right)
+
\left( \begin{array}{ccc}
2\eps_B   & \delta & \delta \\
\delta  & \eps_B   & \eps_B \\
\delta  & \eps_B   & \eps_B \\
\end{array} \right)
\right]
\,,
\label{m}
\eeq
where the mass parameters in this texture are related to the mass
eigenvalues by
\begin{eqnarray}
m &=& m_3/2 \,,
\label{em} \\
\eps_B &=& (m_2 +m_1)/4m \,,
\label{eps} \\
\delta &=&\sqrt{2}\,(m_1-m_2)/4m \,.
\label{delta}
\end{eqnarray}
In our analysis we assume hierarchical neutrinos, but
the case of degenerate neutrinos can be treated
similarly. This would lead to different predictions, since the neutrino
mass scales change, typically with larger rates for
charged-lepton-flavour violation. Hence our results for model (B) are
quite conservative.

\section{Supersymmetric Calculations}

We have calculated the rates for processes violating
charged lepton number in both frameworks,
including complete sets of one-loop sparticle diagrams. We
parametrize the universal soft supersymmetry-breaking masses by
the GUT-scale parameters $m_0$ and $m_{1/2}$, for sfermions and gauginos
respectively, and use the renormalization-group equations of the CMSSM to
calculate the low-energy sparticle masses
\cite{neutrinoLFVs}. Other relevant free parameters
of the MSSM are the trilinear coupling $A$, the sign of the Higgs mixing
parameter $\mu$, and the ratio of Higgs vev's, $\tan\beta$. Here we
consider only $\mu
> 0$, since this is the sign favoured by $g_\mu - 2$. 
The physical charged-slepton masses are found by numerical
diagonalization of the following matrix:
\begin{equation}
\label{eq:66}
\tilde{m}_{\ell}^2=
\left(\begin{array}{cc} m_{LL}^2&m_{LR}^2\\
m_{RL}^2&m_{RR}^2 \end{array}\right)
\end{equation}
where all the entries are $3 \times 3$ matrices in flavour space.
Using the superfield basis where the Yukawa coupling matrix
$\lambda_{\ell}$ is diagonal, we can write:
\begin{eqnarray}
m_{LL}^2&=& (m_{\tilde{\ell}}^\delta)^2 + 
\delta m_{\nu_D}^2+m_{\ell}^2 -
M_Z^2       
(\frac{1}{2} -sin^2\theta_W) \cos 2\beta\\
m_{RR}^2&=& (m_{\tilde{e_R}}^\delta)^2+m_{\ell}^2
 -M_Z^2 \sin^2\theta_W \cos 2\beta \\
m_{RL}^2&=& (A_e^\delta +\delta A_e - \mu \tan\beta) m_{\ell}\\
m_{LR}^2&=& m_{RL}^{2\dagger}
\label{bits}
\end{eqnarray}
where $(m_{\tilde{\ell}}^\delta)^2, (m_{\tilde{e_R}}^\delta)^2$ and
$A_e^\delta$ 
denote the diagonal contributions to the corresponding matrices,
obtained by numerical integration of the
renormalization-group equations, and $\delta m_{\nu_D}^2$ and $\delta A_{l}$
denote the off-diagonal terms that appear because
$\lambda_{\nu_D}$ and $\lambda_{\ell}$ may not be diagonalized
simultaneously - see (\ref{offdiagonal}).
Analogously, for sneutrinos we have
\begin{equation}
{m}_{\tilde{\nu}}^2= (m_{\tilde{\ell}}^\delta)^2+ \delta m_{\nu_D}^2 + \frac{1}{2}
M_Z^2 cos 2\beta
\end{equation}
The mixing parameter $\epsilon$ in (\ref{relation}) is
given in terms of the parameters of this matrix.
In the simplified case where the 
lepton-number-violating mass terms are in the $\tilde{\ell_L}$ sector,
so that the chargino-sneutrino diagram
dominates both $g_\mu - 2$ and $\mu \ra e \gamma$, assuming that the
sparticle masses have (approximately) a common value $\tilde{m}$, 
and making a naive mass-insertion approximation, one
would find
\begin{equation}
\epsilon \approx \frac{(m_{\tilde{\nu}}^2)_{12}}{\tilde{m}^2},
\label{approxepsilon}
\end{equation}
but this is only indicative, and we use complete formulae in our
results below.

We start by fixing the elements of the Yukawa coupling matrices at the GUT
scale to be
consistent with the experimental values of the fermion masses and the  
absolute values of the CKM matrix elements~\cite{GELLN}. This is done by
choosing appropriate coefficients of order one in the entries of
the lepton matrices. In the notation of~\cite{GELLN}, we choose for
model (A):
\begin{equation}
C_{12} \, = \, 0.77, \; \; C_{23} \, = \, 0.79,
\label{modelA}
\end{equation}
and for model (B):
\begin{equation}
C_{12} \, = \, 2.75, \; \; C_{23} \, = \, 1.13,
\label{modelB}
\end{equation}
with the unspecified coefficients taken as unity.
These coefficients do not change significantly for the two
values of $\tan\beta$ considered in the present work.

We then use the full Yukawa coupling matrices in the renormalization-group
equations, including the effects of $\lambda_{\nu_D}$ on the CMMSM
parameters at the see-saw mass scale~\cite{CI}.  We have checked that our
results using the full matricial forms for the Yukawa couplings do not
differ significantly from the common approach of considering diagonal
Yukawa matrices and neglecting the lighter generations.  We check our
results by constructing the slepton mass matrices in the superfield base
where the $\lambda_{\ell}$ are diagonal and inserting the non--diagonal
elements induced by the presence of $\lambda_{\nu_D}$ on the
renormalization-group equations between the GUT and see-saw mass scales. 
Finally, we use the full matricial forms for all the parameters which
appear in the vertices in the diagrams of Fig.~\ref{figure1}, and the
results of~\cite{HMTY} to calculate the rates for $\mu \ra e \gamma$ and
$\mu \ra e$ conversion, and those of~\cite{KAONS} to calculate the rate
for $K^0_L \ra \mu e$.

\section{Constraints on the CMSSM}

We display in the remaining figures the $(m_0, m_{1/2})$ planes in the
CMSSM for $\tan \beta = 10, 30$, assuming $\mu > 0$ as suggested by the
sign of $\delta a_\mu$, and $A_0 = 0$. The experimental constraints on the
CMSSM are taken from~\cite{EFGOSi,ENO}, where further details of their
implementation can be found. We note in particular that the following
choices are used here for the pole mass of the top quark: $m_t = 175$~GeV,
and for the running mass of the bottom quark:  $m_b(m_b)^{\overline
{MS}}_{{SM}} = 4.25$~GeV. We combine the constraints given
in~\cite{EFGOSi,ENO} with the contours suggested by the neutrino mass
textures introduced above for ${\cal B}(\mu \ra e \gamma)$ in
Fig.~\ref{fig:muegamma}, for ${\cal B}(\mu^- Ti \ra e^- Ti)$ in
Fig.~\ref{fig:meco}, and for ${\cal B}( K^0_L \ra \mu^\pm e^\mp)$ in
Fig.~\ref{fig:Kmue}. 

The dark (brick-red) shaded regions in the $(m_0, m_{1/2})$ planes in
these figures are
excluded~\cite{EFGOSi} because the lightest supersymmetric particle is the
lighter
$\tilde \tau$, which is disallowed by the astrophysical requirement that
cold dark matter be electrically neutral. The light (turquoise) shaded
regions are those where the LSP is the lightest neutralino $\chi$, and its
cosmological relic density $\ohsq$ lies in the favoured range $0.1 \le
\ohsq \le
0.3$~\cite{EFGOSi}.
Lower values of $\ohsq$ would be possible if there are other sources
of cold dark matter, whereas larger values of $\ohsq$, which occur
generically at larger values of $m_{1/2}$ and $m_0$, are excluded by
cosmology. 

We display as (red)  dash-dotted lines mass contours for the lightest
CMSSM Higgs boson: $m_h = 113, 117~{\rm GeV}$, as calculated
in~\cite{EFGOSi}. This range corresponds
roughly to values of $m_{1/2}$ and $m_0$ that are compatible, within
theoretical errors, with the LEP Higgs `signal' at $m_H =
115^{+1.3}_{-0.7}$~GeV~\cite{LEPHiggs}, for our default choices of $A_0,
m_t$ and
$m_b(m_b)^{\overline {MS}}_{{SM}}$. There is good overall consistency
between $m_h$ and the other constraints for $10 \lappeq \tan \beta
\lappeq 55$, but we do not insist on the range $113~{\rm GeV} \le m_h
\le 117~{\rm GeV}$, in view of the theoretical uncertainties and because
the LEP Higgs `signal' might turn out to be a false alarm, in which case
$m_h$ could be larger. 

The medium dark (green) shaded regions in panels (b) and (d) are excluded
by our implementation of the $b \to s \gamma$ constraint.  As described
in~\cite{EFGOSi}, we use the latest NLO QCD calculations for large $\tan 
\beta$ and allow values of $m_{1/2}$ and $m_0$ that, after
including the expected theoretical errors due to the scale choice and
model dependences, may fall within the 95\% confidence level range $2.33
\times 10^{-4} < {\cal B}(b \to s < \gamma) < 4.15 \times 10^{-4}$.  In
the panels (a) and (c) of the figures, for $\tan \beta = 10$, there is no
relevant constraint from $b \to s \gamma$, and we display as a dashed line
the LEP lower limit $m_{\chi^\pm} > 104$~GeV. There is a similar
constraint in the panels (b)  and (d), for $\tan \beta = 30$, which is
omitted for clarity.

To complement our summary of~\cite{EFGOSi}, we note that the regions
allowed by the E821 measurement of $a_\mu$ at the 2-$\sigma$
level~\cite{BNLE821} are
shown in the figures as light (pink) shaded regions with solid black
line boundaries~\cite{ENO}.  Also shown as black dashed lines are the
regions favoured by $a_\mu$ at the 1-$\sigma$ level.  We emphasize
the impressive consistency between the constraints from $a_\mu$, $m_h$,
$b \to s \gamma$ and cosmology for $\tan \beta \gappeq 10$. As discussed
in~\cite{ENO}, combining all the other constraints with the 1-$\sigma$
range for $a_\mu$, one finds quite small allowed regions of the
$(m_{1/2}, m_0)$ plane centred on: $\sim (250, 100)$~GeV for $\tan \beta
= 10$ (see panels (a)  and (c))  and $\sim (350, 170)$~GeV for $\tan
\beta = 30$ (see panels (b)  and (d)).

\section{Results for Processes Violating Charged-Lepton Number}

\begin{figure}
\vspace*{-0.75in}
\hspace*{-.35in}
\begin{minipage}{8in}
\epsfig{file=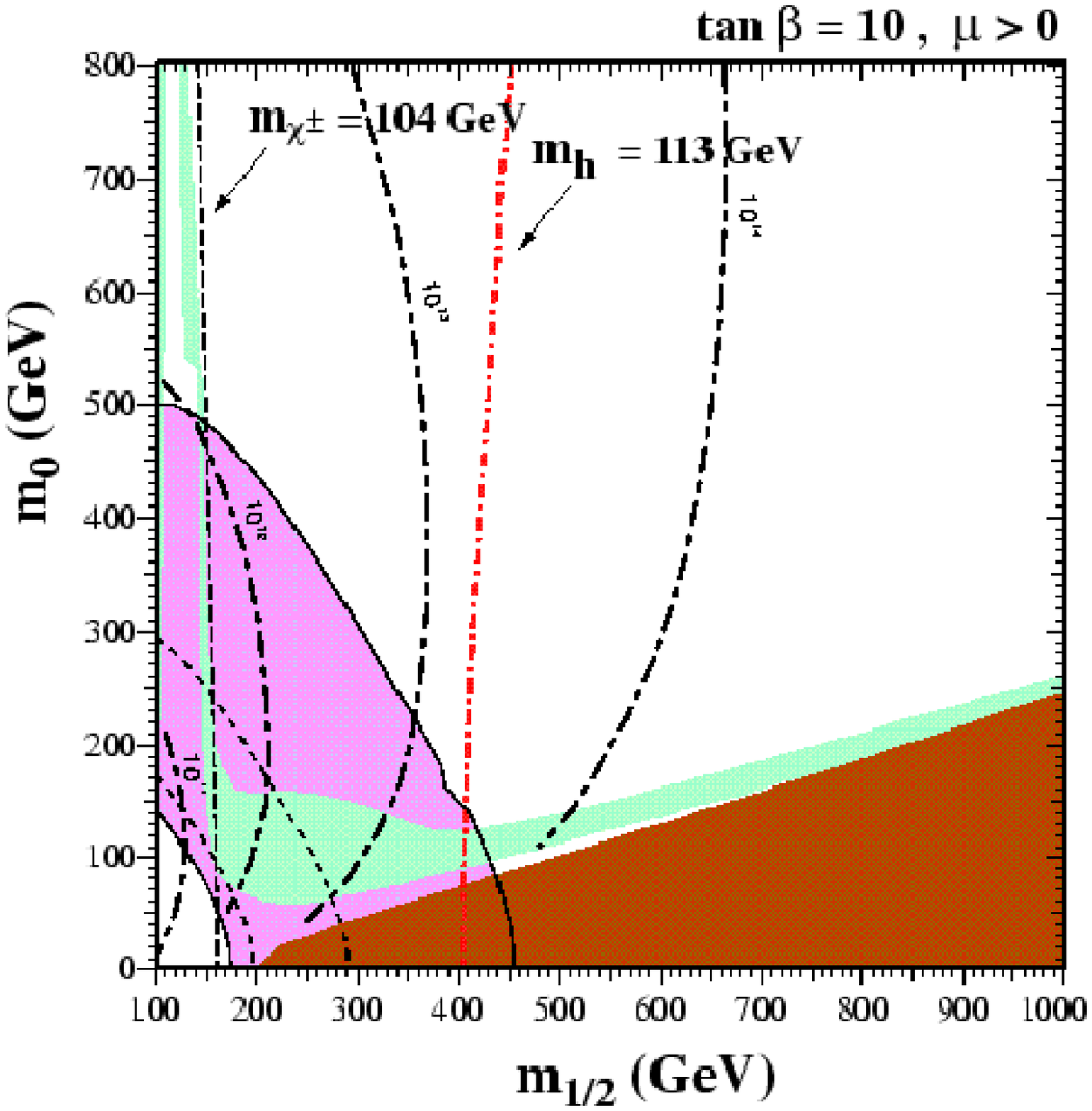,height=3.5in}
\epsfig{file=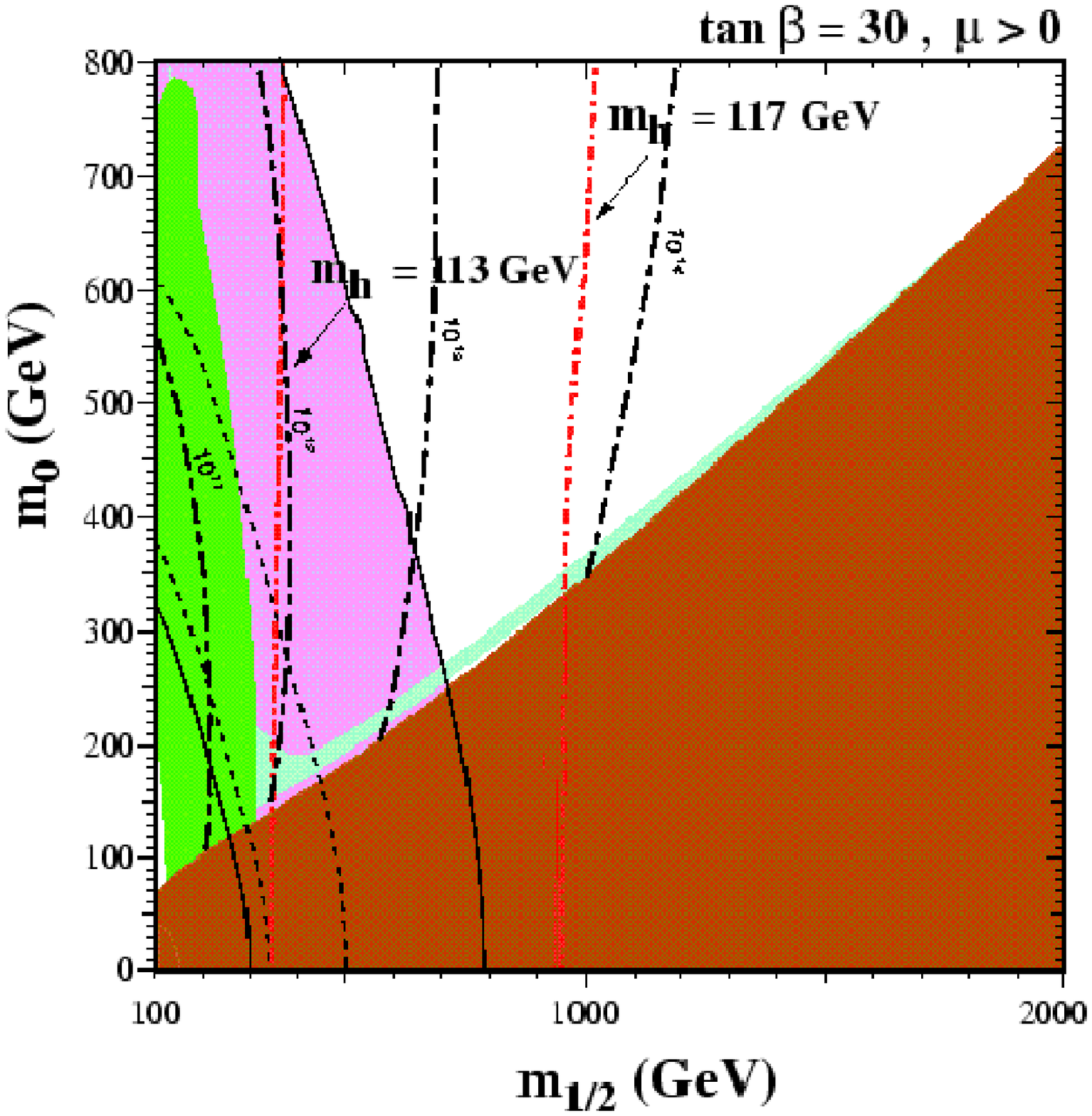,height=3.5in} \hfill
\end{minipage}
\hspace*{-.35in}
\begin{minipage}{8in}
\epsfig{file=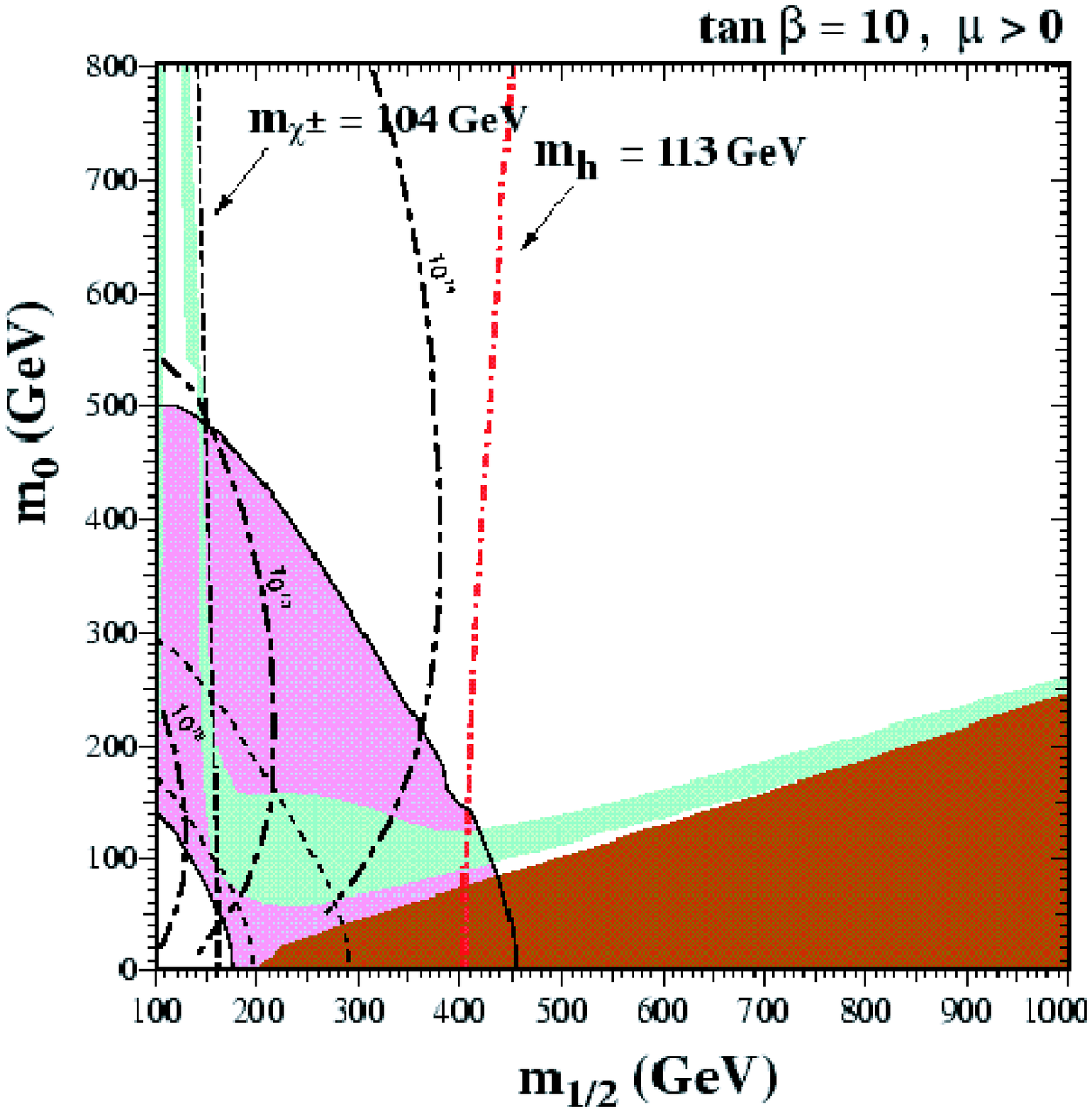,height=3.5in}
\epsfig{file=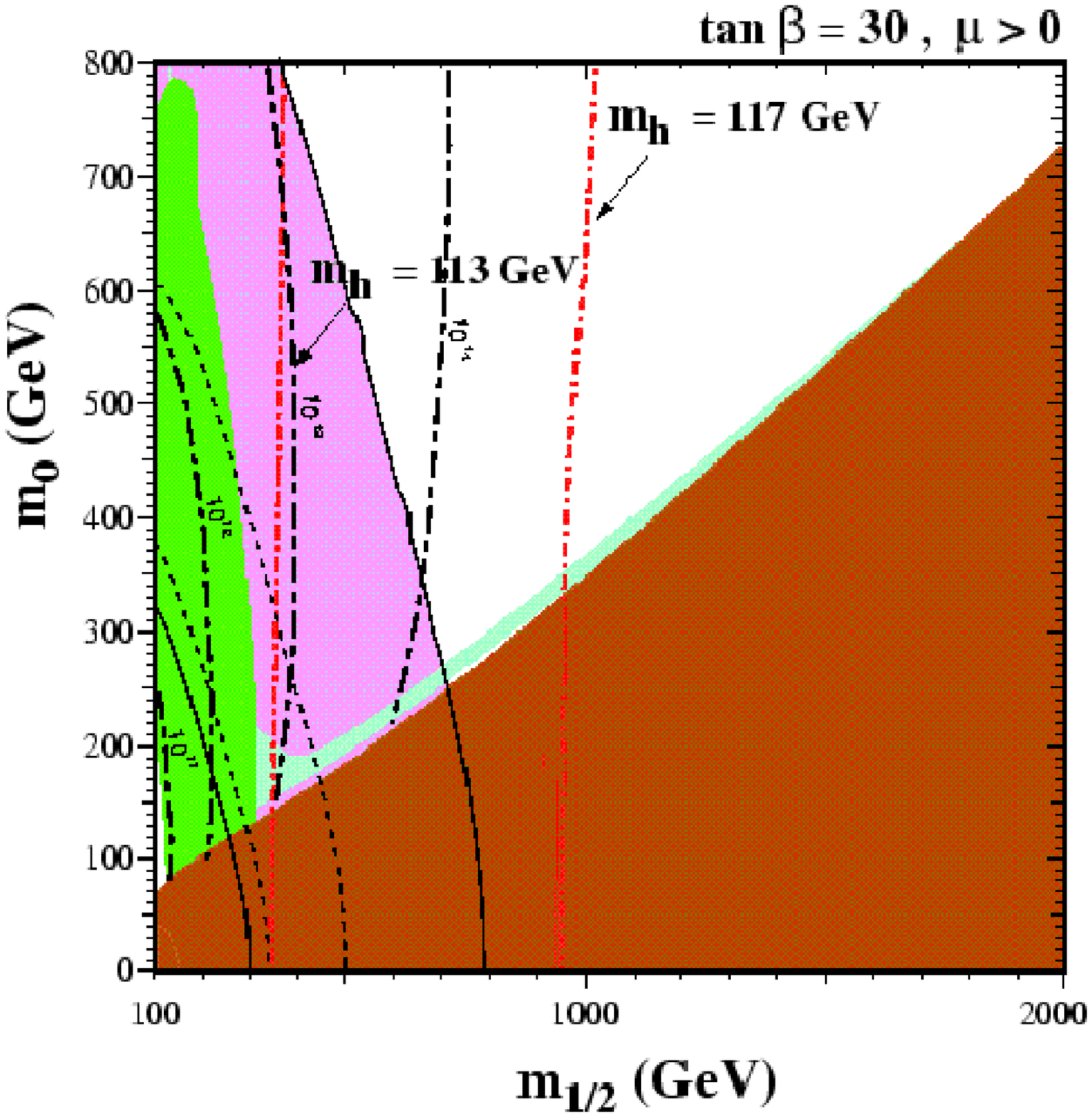,height=3.5in} \hfill
\end{minipage}
\caption{\label{fig:muegamma}
{\it The contours ${\cal B}(\mu \ra e \gamma) = 10^{-11}, 10^{-12},
10^{-13}$ and $10^{-14}$ in (a, b) texture (A) and (c, d) texture (B) are
shown as
dash-dotted black lines in the $(m_{1/2}, m_0)$ planes for $\mu > 0$ and
$\tan \beta =$ (a, c) 10 and (b, d) 30. Other constraints in these planes
are taken from~\cite{EFGOSi}, assuming $A_0 = 0, m_t = 175$~GeV and
$m_b(m_b)^{\overline {MS}}_{SM} = 4.25$~GeV. The regions allowed by the
E821 measurement of $a_\mu$ at the 2-$\sigma$ level~\cite{BNLE821} are
shaded (pink) and
bounded by solid black lines, with dashed lines indicating the 1-$\sigma$
ranges~\cite{ENO}. The dark (brick-red) shaded regions are excluded
because the LSP is the charged ${\tilde \tau}_1$, and the light
(turquoise) shaded regions are those with \protect\mbox{$0.1\leq\ohsq\leq
0.3$} that are preferred by cosmology. We show the contours $m_h = 113,
117$~GeV, and in panels (a, c) the contour $m_{\chi^\pm} = 104$~GeV.  The
medium (dark green) shaded regions are excluded by $b \to s \gamma$.  }}
\end{figure}

Fig.~\ref{fig:muegamma} displays the predictions for ${\cal B}(\mu \ra e
\gamma)$ in texture (A) (panels (a) and (b)) and texture (B) (panels (c)
and (d)). We see that texture (A) generically predicts that ${\cal
B}(\mu \ra e \gamma)$ should occur {\it within one or two (two or three)
orders of magnitude of the present experimental upper limit} if $a_\mu$
lies within the 1(2)-$\sigma$ range suggested by E821~\cite{BNLE821}.
Within this model, the experimental upper limit on ${\cal B}(\mu \ra e
\gamma) < 1.2 \times 10^{-11}$ excludes a domain of the $(m_{1/2}, m_0)$
plane, close to the origin, that may be compared to that
excluded by slepton searches at LEP. When $\tan \beta \sim 10$, it also
has a similar effect to that of the upper limit on the supersymmetric
contribution to $g_\mu - 2$. For larger $\tan \beta \sim 30$, the
constraint due to the present upper limit on ${\cal B}(\mu \ra e
\gamma)$ is intermediate between the $g_\mu - 2$ and $b \ra s \gamma$
constraints. In model (B), we find values of ${\cal B}(\mu \ra e \gamma)$
that are characteristically about an order of magnitude smaller than in
model (A) in the parameter region allowed by cosmology~\cite{EFGOSi} and
$g_\mu - 2$~\cite{ENO}.

\begin{figure}
\vspace*{-0.75in}
\hspace*{-.35in}
\begin{minipage}{8in}
\epsfig{file=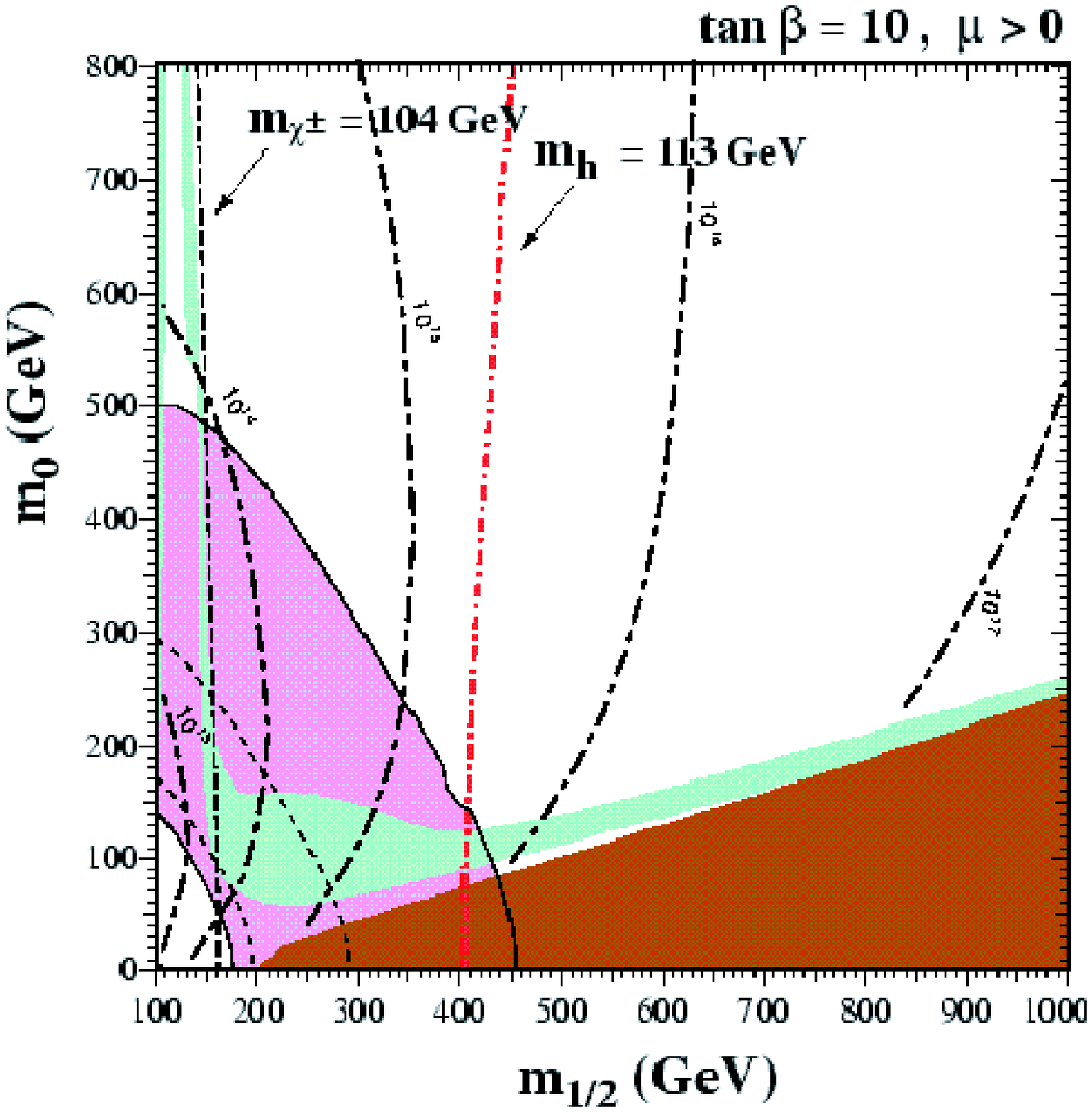,height=3.5in}
\epsfig{file=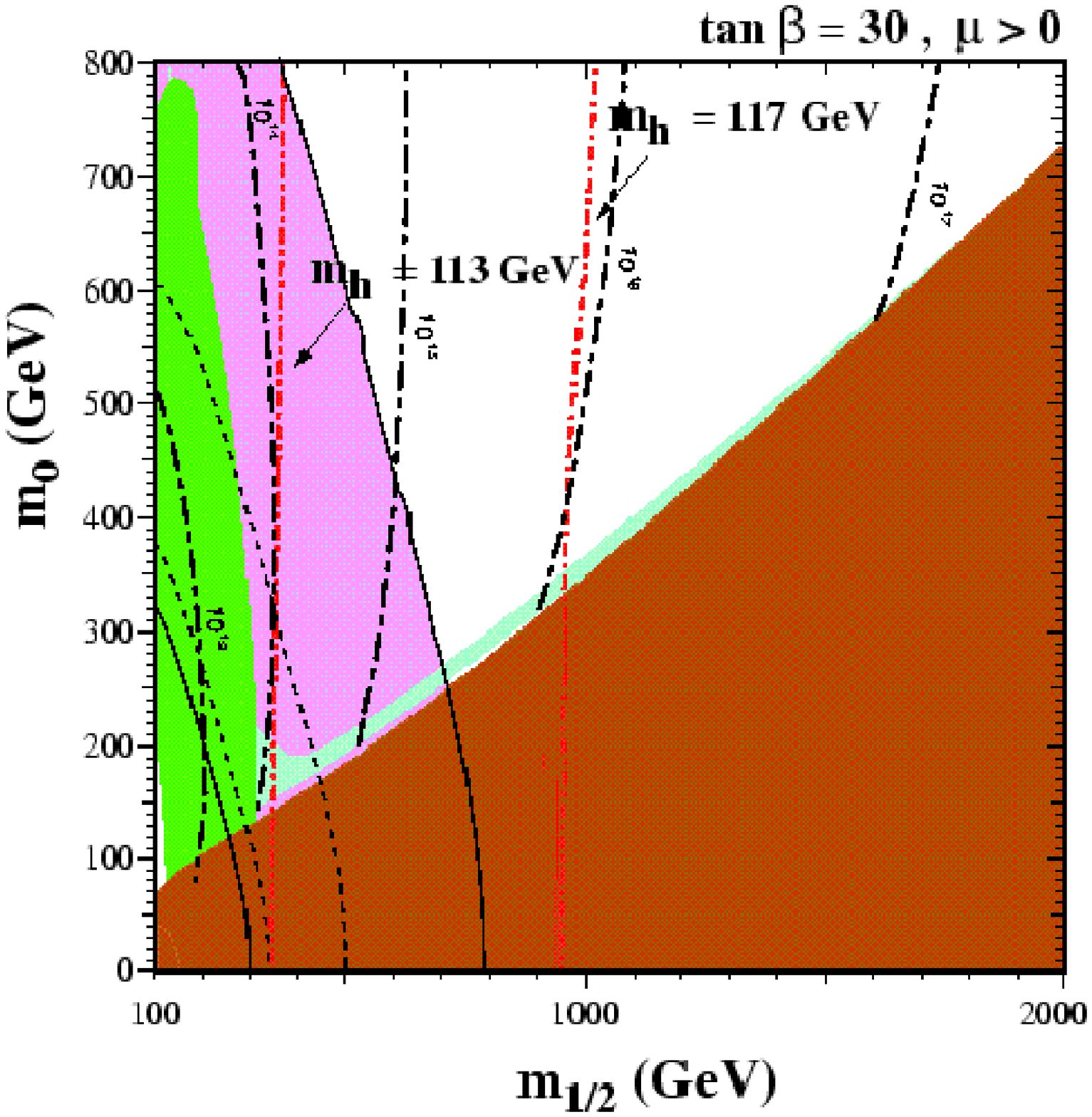,height=3.5in} \hfill
\end{minipage}
\hspace*{-.35in}
\begin{minipage}{8in}
\epsfig{file=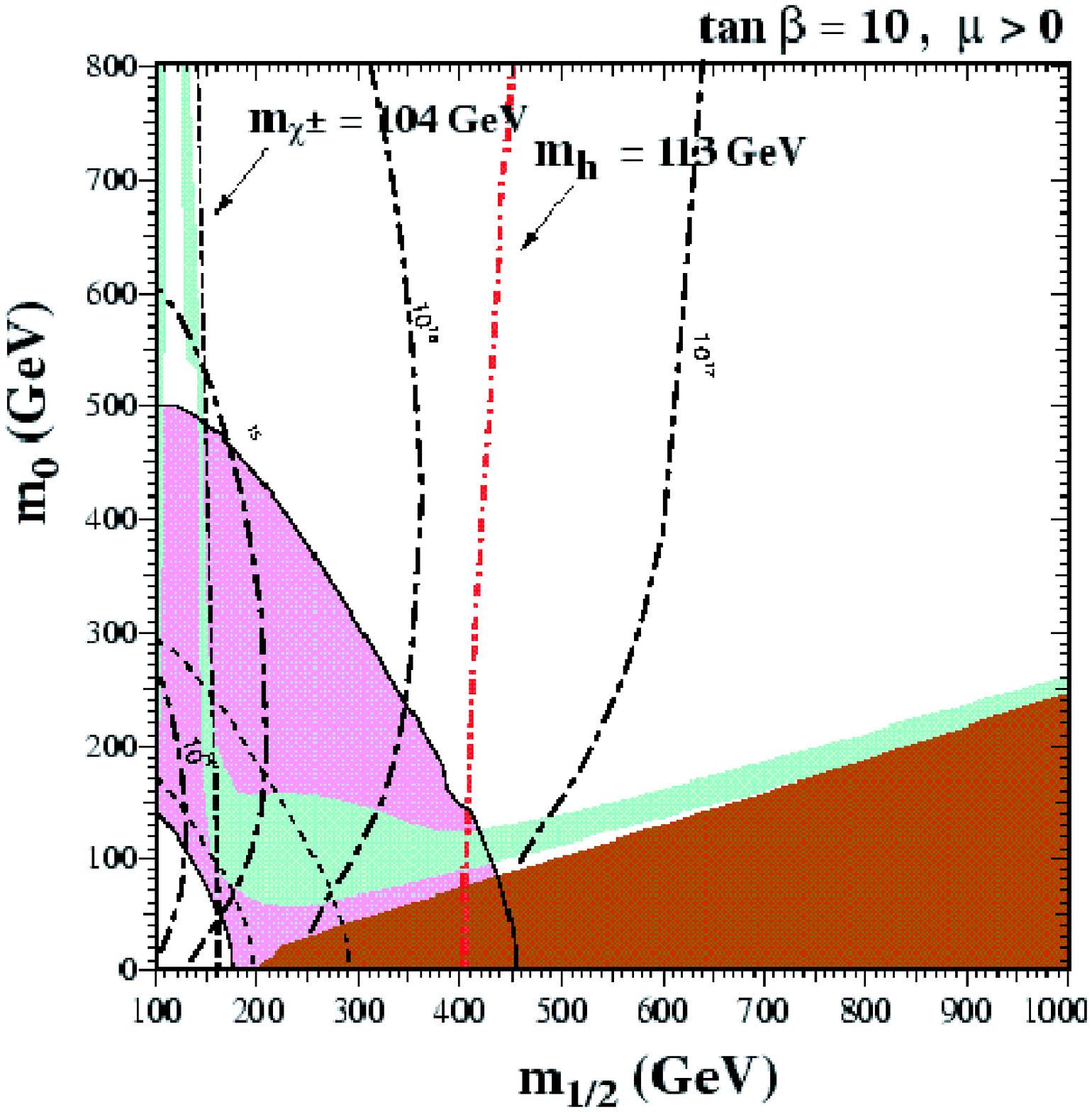,height=3.5in}
\epsfig{file=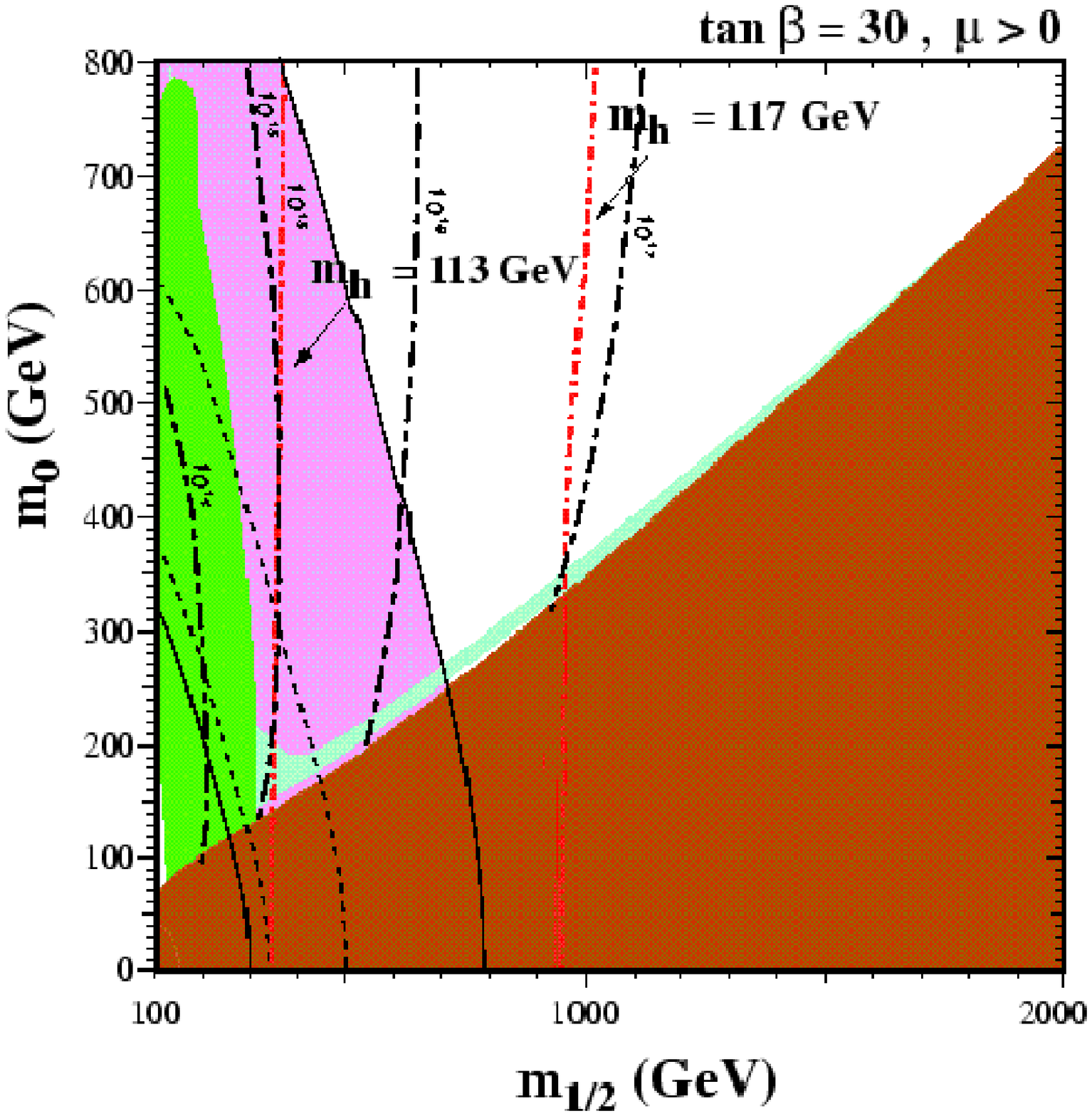,height=3.5in} \hfill
\end{minipage}
\caption{\label{fig:meco}
{\it The contours ${\cal B}(\mu^- Ti \ra e^- Ti) = 10^{-13}, 10^{-14},
10^{-15},
10^{-16}$ and $10^{-17}$ in (a, b) texture (A) and (c, d) texture (B) are
shown as   
dash-dotted black lines in the $(m_{1/2}, m_0)$ planes for $\mu > 0$ and
$\tan \beta =$ (a, c) 10 and (b, d) 30. Other constraints in these planes
are taken from~\cite{EFGOSi,ENO}, as described in the caption of
Fig.~\ref{fig:muegamma}.}}
\end{figure}

As already mentioned, the rate for $\mu^- Ti \ra e^- Ti$ conversion is
linked to that for ${\cal B}(\mu \ra e \gamma)$, with a proportionality
coefficient $\sim 5.6 \times 10^{-3}$ if $\mu^- Ti \ra e^- Ti$ is
dominated by photon exchange. However, this is not necessarily the case,
since $Z^0$ exchange and box diagrams may also contribute, rendering the
ratio ${\cal B}(\mu^- Ti \ra e^- Ti)/ {\cal B}(\mu \ra e \gamma)$
non-universal. Accordingly, we plot in Fig.~\ref{fig:meco} the predictions
of texture (A) (panels (a) and (b))  and texture (B) (panels (c) and (d) 
for ${\cal B}(\mu^- Ti \ra e^- Ti)$.  We see that there are large domains
of the $(m_{1/2}, m_0)$ plane where these textures suggest that ${\cal
B}(\mu^- Ti \ra e^- Ti) > 10^{-16}$, which is the sensitivity of the
proposed MECO experiment~\cite{MECO,KO}.  In model (A), these include all
the regions allowed by $g_\mu - 2$ at the 2-$\sigma$ level, and, in model
(B), most of the allowed region.  {\it We infer that the physics interest
of this proposal is greatly enhanced by the recent result on $g_\mu - 2$
from E821~\cite{BNLE821}.}

\begin{figure}
\vspace*{-0.75in}
\hspace*{-.35in}
\begin{minipage}{8in}
\epsfig{file=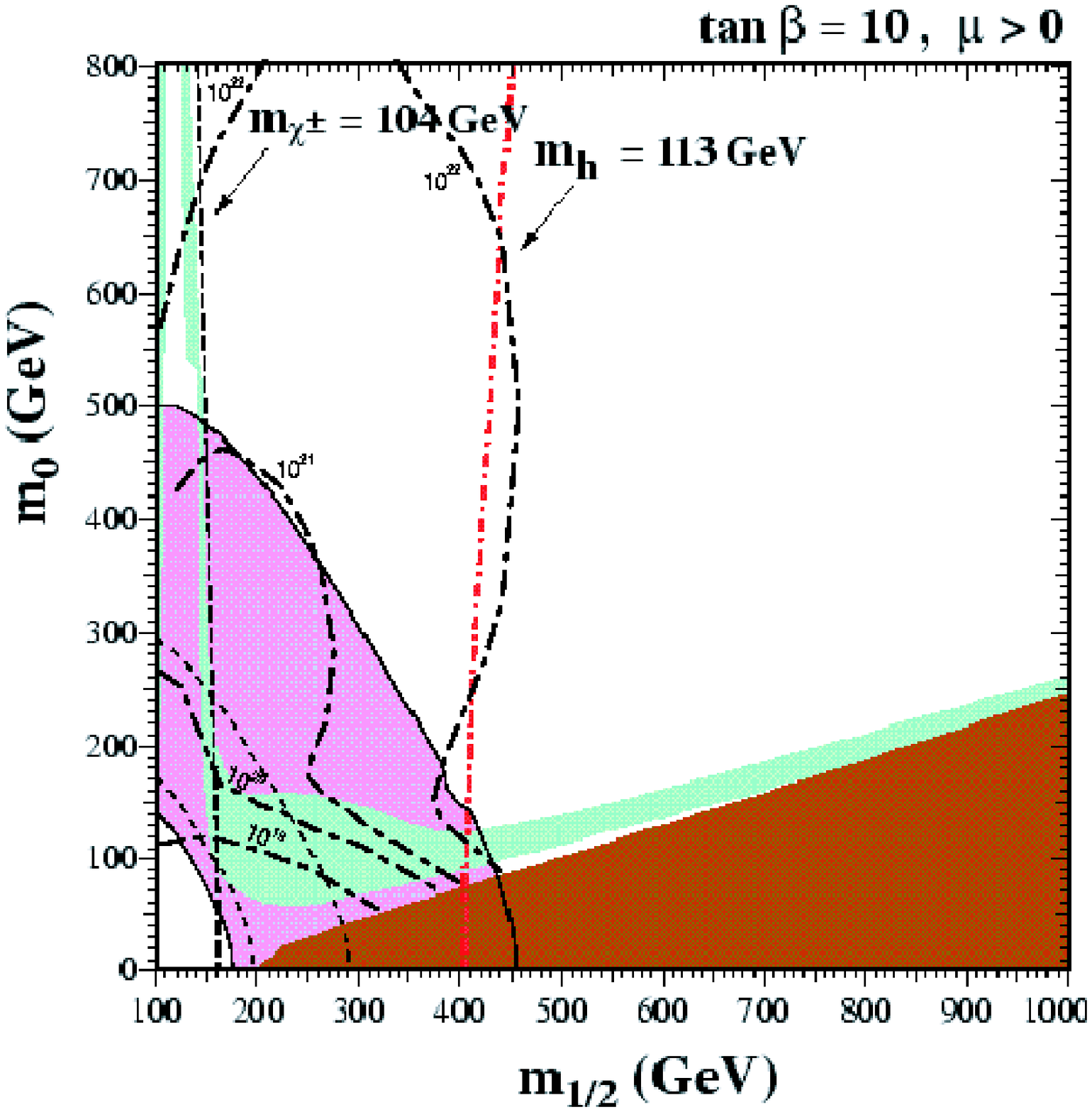,height=3.5in}
\epsfig{file=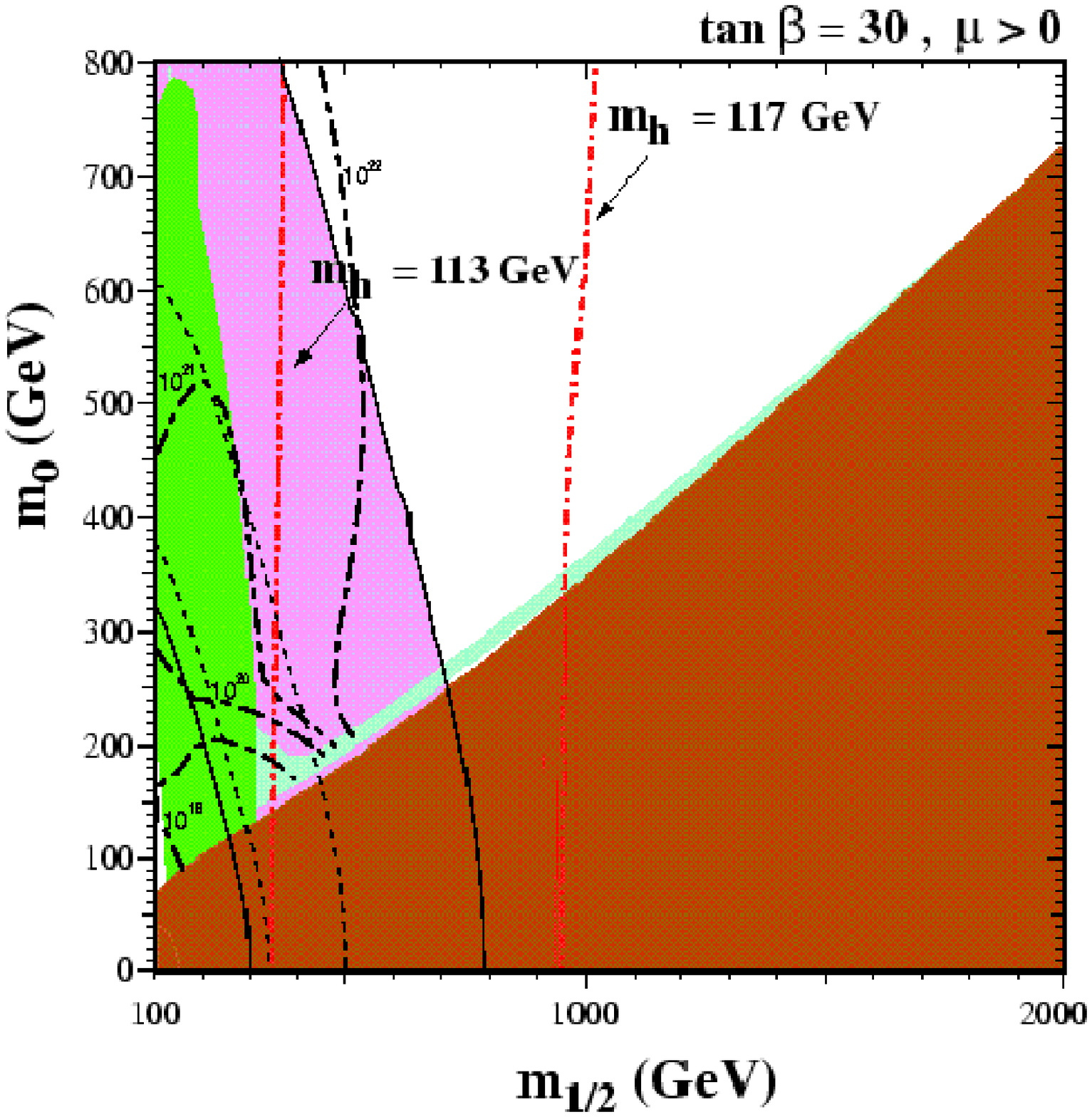,height=3.5in} \hfill
\end{minipage}
\hspace*{-.35in}
\begin{minipage}{8in}
\epsfig{file=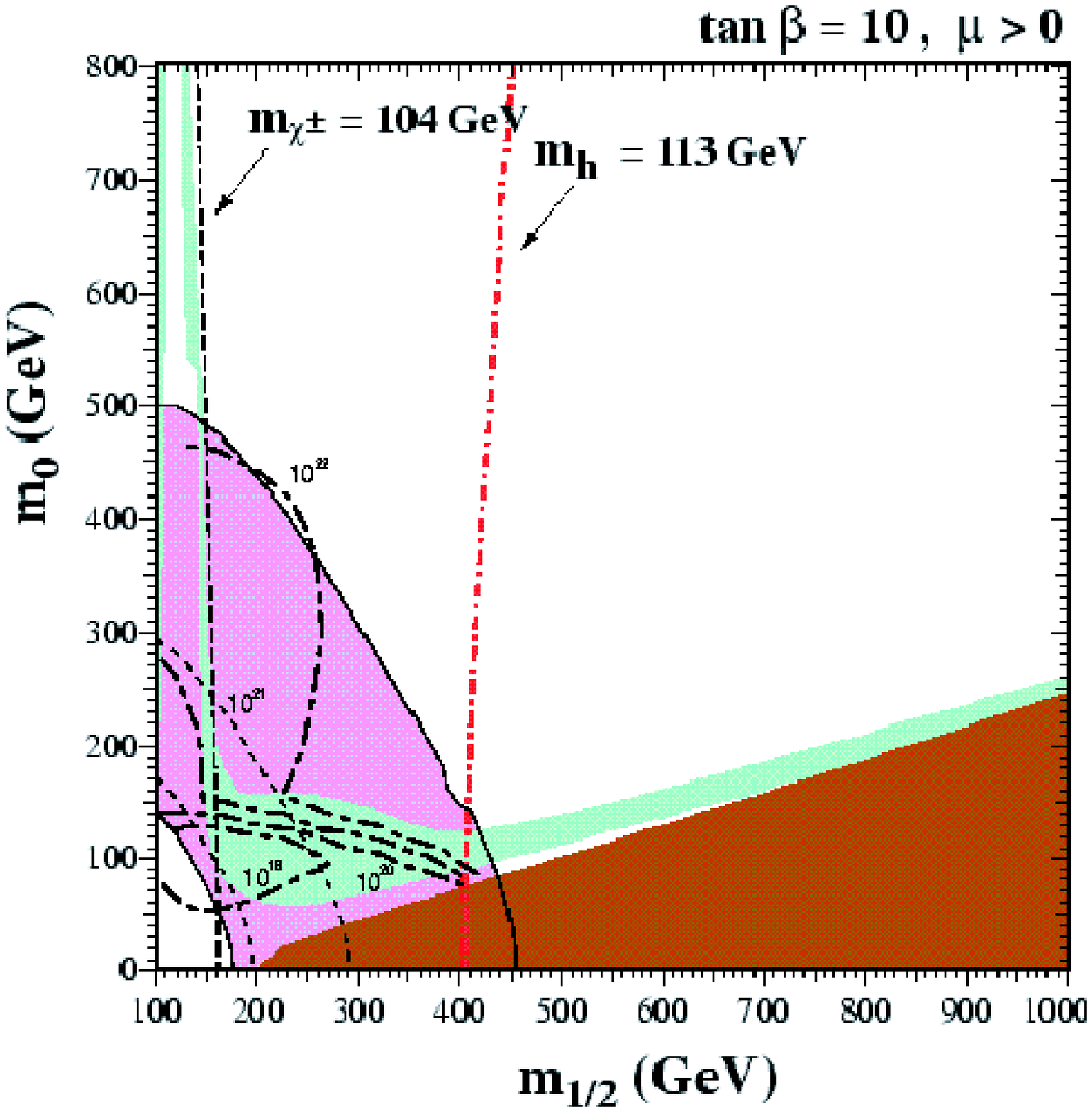,height=3.5in}
\epsfig{file=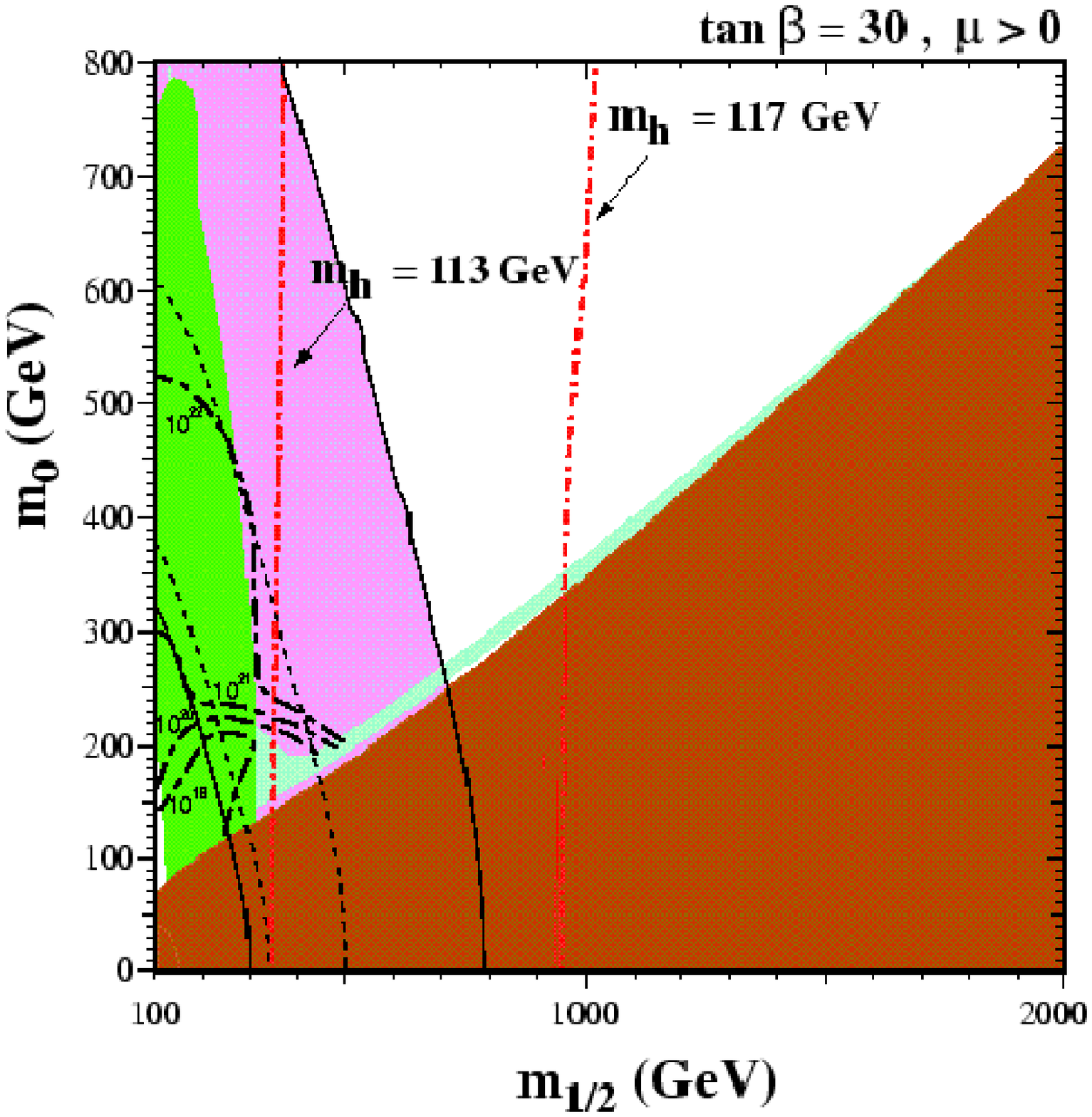,height=3.5in} \hfill
\end{minipage}
\caption{\label{fig:Kmue}
{\it The contours ${\cal B}(K^0_L \ra \mu^\pm e^\mp) = 
10^{-18}, 10^{-20}, 10^{-21}$ and
$10^{-22}$ in (a, b) texture (A) and (c, d) texture (B) are shown as   
dash-dotted black lines in the $(m_{1/2}, m_0)$ planes for $\mu > 0$ and
$\tan \beta =$ (a, c) 10 and (b, d) 30. Other constraints in these planes
are taken from~\cite{EFGOSi,ENO}, as described in the text.}}
\end{figure}

Finally, we plot in Fig.~\ref{fig:Kmue} the corresponding model
predictions for ${\cal B}(K^0_L \ra \mu^\pm e^\mp)$. This process is very
interesting~\cite{KAONS}, because it combines flavour violation in the
quark and lepton sectors. For this same reason, one expects rather small
values of ${\cal B}(K^0_L \ra \mu^\pm e^\mp)$, far below the present
experimental upper limit. However, again for this same reason, there is
clearly even more uncertainty in the predictions for ${\cal B}(K^0_L \ra
\mu^\pm e^\mp)$ than there was already for ${\cal B}(\mu \ra e \gamma)$
and ${\cal B}(\mu^- Ti \ra e^- Ti)$. The sensitivity to the
lepton mass texture is seen clearly by comparing panels (a,
b) and (c, d) of Fig.~\ref{fig:Kmue}. Nevertheless, we note that ${\cal
B}(K^0_L \ra \mu^\pm e^\mp) > 10^{-18}$ in a significant fraction of the
parameter region favoured by $g_\mu - 2$ at the one-$\sigma$ level. Thus,
we think that {\it this process is of potential interest} at an intense
proton source.

\section{Conclusions}

We have argued in this paper that the BNL E821 measurement of $g_\mu -
2$~\cite{BNLE821},
taken at face value, may be used to normalize predictions for the
charged-lepton-number-violating processes $\mu \ra e \gamma, \mu \ra e$
conversion and $K_L^0 \ra \mu^\pm e^\mp$, within a supersymmetric GUT
framework. We have illustrated our argument with a couple of specific
textures for fermion masses that are consistent with the data on neutrino
oscillations. In these examples, we find that $\mu \ra e \gamma$ decay may
appear at a rate within one or two (two or three) orders of magnitude of
the present experimental upper limit if $g_\mu - 2$ lies within its
present one- (two-)$\sigma$ range. These models also make us optimistic
that $\mu \ra e$ conversion on heavy nuclei may be accessible to the next
round of experiments~\cite{MECO,KO}. The prospects for observing $K_L^0 
\ra
\mu^\pm e^\mp$ decay are not so rosy, but this decay might also be
accessible to some future round of experiments with an intense proton
source. 

The discovery of neutrino oscillations has been a major breakthrough in
flavour physics. If confirmed, the deviation of $g_\mu - 2$ from the SM
prediction would be a breakthrough towards new physics at the TeV scale. 
Their combination suggests not only that the conservation of charged
lepton number is not sacred, but also that its violation may soon be
observable. If so, this would be an invaluable new window on the physics
of lepton flavour, as well as on physics at the TeV scale. It would
provide a bridge between neutrino oscillations and accelerator physics,
as well as yield novel information on
lepton mixing. 

{\it We encourage strongly the most sensitive possible
experiments to probe the violation of charged lepton number.}

\section*{Acknowledgements}

The research of D.~F.~C. has been supported by F.C.T. PRAXIS
XXI/BD/9416/96. The research of M.E.G was supported by the European Union
under TMR contract No. ERBFMRX--CT96--009.  We thank Keith Olive and
J.~C.~Rom\~ao for valuable discussions related to this analysis.

\end{document}